\documentclass[a4paper,11pt]{article}

\usepackage{amssymb,amsmath,amsthm}
\usepackage{dsfont}
\usepackage{fullpage}
\usepackage{graphicx}
\usepackage{hyperref}
\usepackage{listings} 
\usepackage{courier} 

\newcommand{\eg}{\emph{eg.}}
\newcommand{\ie}{\emph{ie.}}
\newcommand{\Cplx}{\ensuremath{\mathbb{C}}}
\newcommand{\Mn}[1]{\ensuremath{\mathds{M}_n({#1})}}
\newcommand{\MC}[1]{\ensuremath{\mathds{M}_{#1}(\Cplx)}}

\newcommand{\ket}[1]{\ensuremath{|#1\rangle}}

\newcommand{\Span}[1]{\ensuremath{\mathrm{span}\{#1\}}}

\DeclareMathOperator{\res}{\mathbf{res}}

\DeclareMathOperator{\tr}{\mathbf{tr}}

\newtheorem{qrule}{Rule}
\newtheorem{altqrule}{Alternative Rule}
\newtheorem{qfact}{Fact}

\newcounter{note}
\title{0-th quantization or quantum (information) theory in 42 minutes
\\[12pt]\normalsize{Technical Report IITiS-ZKSI-2013-1}}

\author{Jaros{\l}aw A. Miszczak\\Institute of Theoretical and Applied
Informatics, Polish Academy of Sciences\\Ba{\l}tycka 5, 44-100 Gliwice, Poland}

\date{09/03/2013 (v. 0.17)}

\begin{document}
\maketitle

\begin{flushright}
\begin{minipage}{0.42\textwidth}
\emph{\small Many physicists will tell you that those glorious days of brand new
theories are over. But I don't think so. There are many clues indicating that
one should maintain an open mind. }
\end{minipage}\\[4.2pt]
\small D.M. Greenberger \cite{greenberger12tic}
\end{flushright}

\begin{abstract}
We present a concise introduction of basic concepts in quantum information
theory and quantum mechanics prepared as an introduction for a general audience.
In our approach the rules of quantum mechanics are presented in a simple form of
rules describing the method of constructing quantum objects corresponding to
classical objects. As a byproduct of the introduced approach, we present some
alternative rules and use them to describe basic ingredients of quantum
information theory using different types of objects.
\end{abstract}

\section{Introduction}
Quantum mechanics provides a plethora of phenomena which appear to be
counterintuitive or counterfactual~\cite{jozsa99quantum,hosten06counterfactual,
non09counterfactual} and which are responsible for speed or security boost
offered by quantum information processing. At the same time, these effects,
including quantum superposition and quantum entanglement, are hard to understand
for newcomers in the field~\cite{goff06quantum}.

The main goal of this report is to provide an alternative approach facilitating
the introduction of basic concepts used in the description of quantum systems.
It is done in terms of simple rules for translating classical concepts into
quantum counterparts. We propose to call this approach a \emph{zeroth
quantization}, referring to the standard correspondence between classical and
quantum mechanics described as the \emph{first
quantization}~\cite{dirac81principles}.

Using the introduced approach we describe some alternative methods of
constructing basic ingredients of quantum information theory. We achieve this by
substituting common mapping between classical and quantum information theory,
with alternative rules operating on higher-dimensional objects. 

This report is organized as follows.
In Section~\ref{sec:states} we introduce the concepts of quantum pure states and
the entangled states. 
In Section~\ref{sec:gates} we introduce quantum gates required to process
information encoded in quantum states.
In Section~\ref{sec:alternative} we attempt to provide alternative rules for
the zeroth quantization of states and describe the form of resulting operations.
Finally, in Section~\ref{sec:final} we provide some concluding remarks.

\section{Pure states}\label{sec:states}
We start with the basic ingredient needed in any physical theory, namely the
states. The path chosen here resembles the one used commonly in the
considerations related to the models of computation, where one starts with the
deterministic model (\eg\ deterministic finite automaton) and subsequently
'upgrades' it to include probabilistic or nondeterministic
behavior~\cite{miszczak12high-level}.

In our case we begin with the notion of states used in classical (information)
theory and extend it to the quantum realm. We refer to these states as
\emph{pure} in order to distinguish them from the general states (\ie\
\emph{mixed} states), which arise when we deal with statistical mixtures of pure
states and are described by density matrices.

Let us start with the requirement for the pure states obtained by using the
introduced rules to be elements of a linear space. For a classical bit, the
allowed pure states form a set $\{0,1\}$, and if one aims to add two elements
from this set, the result, \ie\ $a0+b1$, is not a valid state.

For this reason we introduce pure quantum states by mapping classical bits to
vectors. For the sake of simplicity we focus our attention on bits. In this
situation the correspondence between classical and quantum systems is described
by the following rule. 

\begin{qrule}[0-th quantization of pure states]\label{qr:pure-states}
Let us map the states as $0 \mapsto \binom{1}{0}$ and $1 \mapsto \binom{0}{1}$.
\end{qrule}

Here $\binom{a}{b}$ denotes a two-dimensional vector with complex elements.Dirac
notation~\cite{dirac39notation} is commonly used to simplify the notation in
quantum mechanics and quantum information theory.

\begin{qfact}[Dirac notation]
$\ket{0}\equiv\binom{1}{0}$, $\ket{1}\equiv\binom{0}{1}$.
\end{qfact}

Now $\ket{0}$ i $\ket{1}$ are just plain vectors, and thus the state of the
quantum bit can be represented by any combination of these two, namely
\begin{equation}
    x_0\ket{0}+x_1\ket{1},\quad x_0,x_1\in\mathbb{C}.
\end{equation}
This expresses the basic requirement for the quantum theory to be linear.

One may ask why it is necessary to use complex numbers in the above. The reason
for this is that one needs to take into account the states which can be obtained
by using the allowed operations. As we will see in Section~\ref{sec:gates}, the
quantization of classical operations allows the operations which can introduce
complex coefficients. 

Rule \ref{qr:pure-states} is also motivated by another assumption, namely
simplicity. The space of unnormalized states resulting from this rule is
described by $k=4$ real parameters. This agrees with Axiom 2 from
\cite{hardy01quantum}, where quantum mechanics is distinguished from classical
one by the number of real parameters required to specify the state. In quantum
mechanics this number grows like $K=N^2$, instead of $K=N$ in the classical
case, where $N$ denotes the number of distinguishable states.

Using Rule~\ref{qr:pure-states} we can easily obtain states which have no direct
counterparts in the classical theory. For example, the system can be described
by a vector
\begin{equation}
\ket{0}+i\ket{1},
\end{equation}
or if we want for our states to be normalized, by a vector
\begin{equation}
\frac{1}{\sqrt{2}}(\ket{0}+i\ket{1}).
\end{equation}
Such states -- superpositions of base states -- are crucial for quantum
algorithms as they allow processing multiply values in a single computational
step. This is exploited in the most prominent way in Grover's algorithm
\cite{grover97haystack}.

\subsection{Compound systems}\label{sec:compound}
In quantum mechanics the most interesting things happen when we use more than
one system or a system with two subsystems. In the case of classical systems the
state of the compound system is described by the state space constructed as a
Cartesian product of spaces used to describe subsystems. In quantum case the
situation is more involved. The state space of the compound system is
constructed as a tensor product of the spaces used to describe the subsystems.

\subsubsection{Two subsystems}
Let us assume for a start that we deal with the system composed of two
subsystems. In this situation we use the following rule.

\begin{qrule}[0-th quantization of a bipartite system]\label{qr:compound-states}
If $(b_0,b_1)\in\{0,1\}\times\{0,1\}$ we put $(b_0,b_1)\mapsto
\ket{b_0}\otimes\ket{b_1}$,
where $\otimes$ denotes the tensor product.
\end{qrule}

The first question appearing here is what exactly the tensor product is. 

For the physicists this term is related mostly to the theory of General
Relativity (see \eg~\cite{schutz}), but in this context tensor is used as a
multi-linear form (map) on $V^p\times(V^\star)^q$ into $F$, where $V$ is some
vector space over $F$. For example, if we get $p=1$ and $q=1$, the tensors are
just matrices. For $p>1$ or $q>1$ the tensors are represented by $n$-way arrays,
which are called tensors as well.

Tensors (multilinear maps) form a linear space and one can define a tensor
product of such spaces. The concept of a tensor product, however, can be defined
for any vector space. The crucial part of this extension is that the tensor
product transforms multilinear maps into linear ones.

\begin{qfact}[Universal property of tensor product] 
If $V$ and $W$ are finite-dimensional vector spaces, then a tensor product of
$V$ and $W$ is a vector space $T$ with a bilinear map $\tau:V\times W\mapsto T$,
such that
\begin{equation}
\forall_{f:V\times W\mapsto X} \exists!_{g:T\mapsto X} f = \tau \circ g,
\end{equation}
\ie\ for any bilinear map $f:V\times W\mapsto X$ into some vector space $X$,
there exists a unique $g:T\mapsto X$, such that $ f = \tau \circ g $.
\end{qfact}

The above property defines a tensor product in a unique way. In the case of
finite-dimensional spaces the tensor product is known as a Kronecker product and
can be found in various areas of applied mathematics~\cite{vanloan00ubiquitous}.

One should note that the use of a tensor product for representing the state of
compound system can be justified by using a different
approach~\cite{aerts78physical}. It is also worth noting that by using a
Kronecker product for describing compound systems, we obtain a theory which
agrees with Axiom 4 from \cite{hardy01quantum}, which requires from the system
composed of two subsystems with dimensions $N_A$ and $N_B$ to be described by a
space of dimension $N_A N_B$.

In the case of two quantum bits, the base states (\ie\ states obtained directly
by the application of Rule~\ref{qr:compound-states}) are
\begin{equation}
\ket{00} \equiv \ket{0}\otimes\ket{0},\ 
\ket{01}\equiv \ket{0}\otimes\ket{1},\ 
\ket{10}\equiv \ket{1}\otimes\ket{0},\ 
\ket{11}\equiv \ket{1}\otimes\ket{1},
\end{equation}
and the system can be also in any state obtained as a linear combination of the
above states
\begin{equation}
\ket{x}=x_{00} \ket{00} + x_{01} \ket{01} + x_{10} \ket{10} + x_{11} \ket{11}.
\end{equation}

In the compound systems we can distinguish between separable and non-separable
(\ie\ entangled) states. This distinction is based on the following fact, known
as the Schmidt decomposition~\cite{schmidt07zur}.

\begin{qfact}[Schmidt decomposition]
The matrix of coefficients 
\begin{equation}
M_{\ket{x}}=\left(\begin{matrix}
x_{00} & x_{01}\\
x_{10} & x_{11}
\end{matrix}
\right)
\end{equation}
can be represented in a diagonal form by local changes of bases on both
subsystems.
\end{qfact}

The representation mentioned above is obtained by calculating a Singular Value
Decomposition of the coefficient matrix and can be used for any nitary
space~\cite{miszczak11singular}.

Using the Schmidt decomposition we can introduce two different classes of
states, namely
\begin{itemize}
    \item separable states -- vectors such that the diagonalized matrix has only
    one entry,
    \item entangled states -- vectors such that the diagonalized matrix has
    two entries.
\end{itemize}

An example of the entangled state is given by so-called Bell state, one of
which reads
\begin{equation}
\ket{\psi^+}=\frac{1}{\sqrt{2}}(\ket{10}+\ket{01}).
\end{equation}
In this case the matrix of coefficients reads
\begin{equation}
M_{\ket{\psi^+}}=\frac{1}{\sqrt{2}}\left(\begin{matrix}
0 & 1\\
1 & 0
\end{matrix}
\right),
\end{equation}
and it has two singular values.

\subsubsection{Three or more subsystems}
The construction of the states describing systems composed of more than two
subsystems relies on a general construction of a tensor product.

\begin{qrule}[0-th quantization of multipartite states]
If $(b_0,b_1,\dots,b_{n-1})\in\{0,1\}^n$ we map this state as
$(b_0,b_1,\dots,b_{n-1})\mapsto \ket{b_0}\otimes\ket{b_1}\otimes \dots
\otimes\ket{b_{n-1}}$.
\end{qrule}

In this situation the construction of the tensor product assures that any
$n$-linear map is translated into a linear map. However, in this situation one
cannot use Schmidt decomposition to distinguish between separable and entangled
states~\cite{peres95higher}. 

Even for the case of three subsystem the situation is far more complicated than
in the case of two subsystems. A classification of states with respect of their
degree of entanglement in three-partite systems is given
in~\cite{acin00generalized}.

\section{Quantum gates}\label{sec:gates}
Having introduced the states we wish to operate on, it is easy to provide a
quantization of allowed operations. Once again, the main rule is to use
linearity of quantum theory. Moreover, it is required from the resulting
operation to be reversible.

\subsection{One-qubit gates}
We start with operations acting on one bit. In the classical case we have only
two allowed reversible operations: identity and negation.

\begin{qrule}[0-th quantization of one-bit circuits]\label{qr:basic-circuits}
For a give reversible classical operation, its quantum counterpart is defined by
its application on the base states $\ket{0}$ and $\ket{1}$.
\end{qrule}

For operations acting on one bit, the simplest examples is the binary negation
(NOT) operation. Its quantum counterpart is give by a matrix which simply
interchanges the base states, 
\begin{equation}
NOT=\left(\begin{matrix}
0 & 1\\
1 & 0
\end{matrix}
\right).
\end{equation}

Now, as we are able to use linear combinations of states, we would like to use
gates which can result in such superpositions. The simplest examples is the
$\sqrt{NOT}$ gate, defined as
\begin{equation}
\sqrt{NOT}=\frac{1}{2}\left(
\begin{matrix}
 i+1 & i-1 \\
 i-1 & i+1 \\
\end{matrix}
\right),
\end{equation}
which results in states half way between the input state and the negated input
state. Gate $\sqrt{NOT}$ acts on the base states $\ket{0}$ and $\ket{1}$ as
\begin{equation}
  \begin{split}
	\sqrt{NOT}\ket{0} &= \frac{1+i}{2}\ket{0} + \frac{1-i}{2}\ket{1},\\
	\sqrt{NOT}\ket{0} &= \frac{1-i}{2}\ket{0} + \frac{1+i}{2}\ket{1}.
  \end{split}
\end{equation}
The use of gates such as $\sqrt{NOT}$ justifies to some degree the need
for using complex numbers in quantum mechanics. 

Although we require any quantum circuit to be reversible, we are able to
introduce the operations which do not have classical counterparts. The most
prominent example is the Hadamard matrix, 
\begin{equation}
H=\frac{1}{2}\left(\begin{matrix}
1 & 1\\
1 & -1
\end{matrix}
\right).
\end{equation}
Another example, this time based on the ability to use complex numbers, is
\begin{equation}\label{eqn:phase-gate}
R(\phi)=\left(\begin{matrix}
1 & 0\\
0 & e^{i\phi}
\end{matrix}
\right).
\end{equation}

The ability of using continuous transformations between the states is the
requirement which distinguishes quantum mechanics from the classical
theory~\cite{hardy01quantum}. Without this requirement, one would be able to
enumerate allowed states of the system and the obtained space of states would be
equivalent to the space of classical states.

\subsection{Two-qubit gates}
The natural extension of Rule~\ref{qr:basic-circuits} provides a method for
constructing quantum gates for reversible 2-bit operations.

\begin{qrule}[0-th quantization of reversible two-bit circuits]
  For a given reversible classical operation acting on two bits,
  $f_2:\{0,1\}\times\{0,1\} \mapsto \{0,1\}\times \{0,1\}.$, its quantum
  counterpart is defined by the application on the base states,
  \begin{equation}
	G_{f_2} \ket{x_1}\ket{x_2} = \ket{f_2(x_1,x_2)}.
  \end{equation}
\end{qrule}

Moreover, the ability to operate on two bits allows introducing a wider class of
operations. In particular, one can introduce irreversible one-bit operations
using the following rule.

\begin{qrule}[0-th quantization of irreversible one-bit
  circuits]\label{qr:one-bit-irreversible}
  For a given irreversible classical operation acting on one bit, $f_1:\{0,1\}
  \mapsto \{0,1\}$, its quantum counterpart is defined as
  \begin{equation}
	G_{f_1}\ket{x_1}\ket{x_2} = \ket{x_1}\ket{f_1(x_1) + x_2 \pmod 2}.
  \end{equation}
\end{qrule}

For example, a quantum gate for binary function $f_1^{(0)}(x)=0$ (\ie\ reset
operation) defined using this rules is given by identity matrix, while
quantum counterpart of binary function $f_1^{(1)}(x)=1$ is given by
\begin{equation}
  \left(
  \begin{smallmatrix}
	0 & 1 & 0 & 0\\
	1 & 0 & 0 & 0\\
	0 & 0 & 0 & 1\\
	0 & 0 & 1 & 0\\
  \end{smallmatrix}
  \right).
\end{equation}

The ability to operate on two bits allows the introduction of the basic
ingredient used in all programming languages, namely conditional statements. The
simplest gate we can consider here is $NOT$, and this gives the classical
pseudocode illustrating conditional statements presented in Listing
\ref{lst:classical-conds}

\begin{lstlisting}[label=lst:classical-conds,caption=Pseudocode for the simplest conditional statement with '\lstinline{==}' denoting comparison and '\lstinline{:=}' denoting assigenemt.]
if $x_1$ == 1 then
  $x_2$ := not($x_2$)
end
\end{lstlisting}

The quantum counterpart of this operation can be obtained by writing the truth
table for this operation. As a result we get the quantum version of the code
from Listing \ref{lst:classical-conds},
\begin{equation}
\begin{split}
  \ket{0}\ket{0} \mapsto \ket{0}\ket{0},\\
  \ket{0}\ket{1} \mapsto \ket{0}\ket{1},\\
  \ket{1}\ket{0} \mapsto \ket{1}\ket{1},\\
  \ket{1}\ket{1} \mapsto \ket{1}\ket{0},
  \end{split}
\end{equation}
and one can easily see that this operation is realized by the gate
\begin{equation}
CNOT = \left(
\begin{smallmatrix}
 1 & 0 & 0 & 0 \\
 0 & 1 & 0 & 0 \\
 0 & 0 & 0 & 1 \\
 0 & 0 & 1 & 0 \\
\end{smallmatrix}
\right).
\end{equation}
The conditional application of a general gate 
\begin{equation}
  U=\left(
  \begin{smallmatrix}
	u_{00} & u_{01} \\
	u_{10} & u_{11} 
  \end{smallmatrix}
  \right),
\end{equation}
results in a conditional operation give as
\begin{equation}
\left(
\begin{smallmatrix}
 1 & 0 & 0 & 0 \\
 0 & 1 & 0 & 0 \\
 0 & 0 & u_{00} & u_{01} \\
 0 & 0 & u_{10} & u_{11} \\
\end{smallmatrix}
\right).
\end{equation}

However, the syntax similar to the pseudo-code in Listing
\ref{lst:classical-conds} is available in some quantum programming languages
\cite{oemerPhD,miszczak12high-level}.

Rule~\ref{qr:one-bit-irreversible} suggests the general method for obtaining
quantum version of general binary functions $f_{m,n}: \{0,1\}^m \mapsto
\{0,1\}^n$.

\begin{qrule}[0-th quantization of irreversible circuits]\label{qr:irreversible}
  For a given irreversible classical operation acting on $m$ bits, $f:\{0,1\}^m
  \mapsto \{0,1\}^n$, its quantum counterpart is defined as
  \begin{equation}
	G_{f}\ket{x}\ket{y} = \ket{x}\ket{f(x_1) + y_1 \pmod 2},
  \end{equation}
  where $\ket{x}$ and $\ket{y}$ denote $m$-qubit and $n$-qubit states. 
\end{qrule}

\section{Alternative rules}\label{sec:alternative}
The rules presented in Sections~\ref{sec:states}-\ref{sec:gates} allow the
reproduction of a standard presentation of quantum information
theory~\cite{NC00}. There is no reason, however, why we cannot try to describe
the quantization of classical information theory using a different set of rules.

\subsection{Zeroth quantization with qutrits and ququarts}
The simplest example of generalizing the standard zeroth quantization is to use
larger space in the place of $\Cplx^2$.

To do this we start by substituting Rule~\ref{qr:pure-states}, with the
following one.

\begin{altqrule}[0th quantization with qutrits]\label{qr:states-3d}
Let us map the states $0$ and $1$ as $0\mapsto (1,0,0)^T$ and $1\mapsto
(0,0,1)^T$.
\end{altqrule}

Using the nomenclature of quantum information theory the above rule maps bit to
qutrit (instead of qubit). However, as we do not use full $\Cplx^3$, the third
of the vectors has to be fixed, \ie\ set to $(0,1,0)^T$.

It is easy to see that in this case $NOT$ operation (\ie\ operation exchanging
the base vectors encoding our bit) reads
\begin{equation}
\left(
\begin{smallmatrix}
 0 & 0 & 1 \\
 0 & 1 & 0 \\
 1 & 0 & 0 \\
\end{smallmatrix}
\right).
\end{equation}
One can also easily obtain a $\sqrt{NOT}$ operation, which, for the above
mapping reads
\begin{equation}
\frac{1}{2}\left(
\begin{smallmatrix}
 1+i & 0 & 1-i \\
 0 & 2 & 0 \\
 1-i & 0 & 1+i \\
\end{smallmatrix}
\right).
\end{equation}

Clearly the above mapping is equivalent to operating on a fixed subspace of the
$\Cplx^3$ -- the vectors from the subspace $\Span{(0,1,0)^T}$ are simply
ignored. Instead, one could try to use a different mapping for $0$, \eg
\begin{equation}
0\mapsto \Span{(1,0,0)^T, (0,1,0)^T},
\end{equation}
but this could not be used to obtain a unitary matrix in place of $NOT$
operation.

The above considerations suggest to use the mapping onto $\Cplx^4$ given in the
following rule.
\begin{altqrule}\label{qr:states-4d}
Let us map the states $0$ and $1$ as 
\begin{equation}
0\mapsto \Span{(1,0,0,0)^T, (0,0,0,1)^T}
\end{equation}
and 
\begin{equation}
1\mapsto
\Span{(0,1,0,0)^T, (0,0,1,0)^T}.
\end{equation}
\end{altqrule}
This basically encodes bits into two-dimensional subspaces of $\Cplx^4$.
Gate $NOT$ obtained by using this mapping can be expressed as
\begin{equation}
NOT_{4}=\left(
\begin{smallmatrix}
 0 & 1 & 0 & 0 \\
 1 & 0 & 0 & 0 \\
 0 & 0 & 0 & 1 \\
 0 & 0 & 1 & 0 \\
\end{smallmatrix}
\right),
\end{equation}
which simply flips $(1,0,0,0)^T\leftrightarrows(0,1,0,0)^T$ and
$(0,0,0,1)^T\leftrightarrows(0,0,1,0)^T$, and results in $\sqrt{NOT}$ operation
given as
\begin{equation}
\frac{1}{2}\left(
\begin{smallmatrix}
 1+i & 1-i & 0 & 0 \\
 1-i & 1+i & 0 & 0 \\
 0 & 0 & 1+i & 1-i \\
 0 & 0 & 1-i & 1+i \\
\end{smallmatrix}
\right).
\end{equation}

However, there are also three other matrices which can be used in place of $NOT$
gate, namely the matrices
\begin{equation}
\left(
\begin{smallmatrix}
 0 & 0 & 1 & 0 \\
 0 & 0 & 0 & 0 \\
 1 & 0 & 0 & 0 \\
 0 & 1 & 0 & 1 \\
\end{smallmatrix}
\right),
\left(
\begin{smallmatrix}
 0 & 0 & 1 & 0 \\
 1 & 0 & 0 & 0 \\
 0 & 0 & 0 & 1 \\
 0 & 1 & 0 & 0 \\
\end{smallmatrix}
\right),
\left(
\begin{smallmatrix}
 0 & 0 & 1 & 0 \\
 0 & 0 & 0 & 1 \\
 1 & 0 & 0 & 0 \\
 0 & 1 & 0 & 0 \\
\end{smallmatrix}
\right).
\end{equation}

Using Alternative Rule~\ref{qr:states-4d} on can also easily construct
$\frac{NOT_4}{2}$ gate, which negates only some bits. Let us assume that this
gate flips $(0 ,0,0,1)^T\leftrightarrows(0,0,1,0)^T$ only. Then it is given as
\begin{equation}
\frac{NOT_4}{2} =
\left(
\begin{smallmatrix}
 1 & 0 & 0 & 0 \\
 0 & 1 & 0 & 0 \\
 0 & 0 & 0 & 1 \\
 0 & 0 & 1 & 0 \\
\end{smallmatrix}
\right),
\end{equation}
which is simply $CNOT$ gate on qubits with control on the first and target on
the second qubit. Gate flipping $(1,0,0,0)^T\leftrightarrows(0,1,0,0)^T$ only is
equivalent to the controlled $NOT$ operation of qubits with the first qubit
being the and target.

\subsection{Zeroth quantization with matrices}\label{sec:alt-matrices}
Taking into account our requirement for the obtained structure to be linear, one
can use any mapping which turns bits into elements of some vector space. 

The most straightforward generalization of the above examples is obtained by
using matrices in the place of vectors.
\begin{altqrule}\label{qr:states-mtx}
Let us map the states $0$ and $1$ as 
\begin{equation}
\begin{split}
0\mapsto \Span{
\left(
    \begin{smallmatrix}
    1 & 0\\
    0 & 0
    \end{smallmatrix}
    \right), 
    \left(\begin{smallmatrix}
    0 & 0\\
    0 & 1
    \end{smallmatrix}
    \right)},\\
    1\mapsto \Span{
    \left(
    \begin{smallmatrix}
    0 & 1\\
    0 & 0
    \end{smallmatrix}
    \right), 
    \left(\begin{smallmatrix}
    0 & 0\\
    1 & 0
    \end{smallmatrix}
    \right)},
	\end{split}
  \end{equation}
\end{altqrule}

In order to construct quantum gates acting on such states, one can note that the
matrix-vector multiplication is composed of dot (\ie\ scalar) products between
vectors forming a matrix and the vector the matrix acts on. The resulting vector
is subsequently transposed. This suggests that in the case of matrices one
should use Hilbert-Schmidt scalar product to define quantum gates. This appears
to be more tricky, but one can easily obtain the form of the basic operations by
using isomorphism between $\Mn{\Cplx}$ and $\Cplx^{n^2}$.

To achieve this we use $\res$ operation~\cite{miszczak11singular}, which maps
elements of $\Mn{\Cplx}$ into vectors using row-order. For examples, if
\begin{equation}
A=\left(
\begin{smallmatrix}
 a_{1,1} & a_{1,2} \\
 a_{2,1} & a_{2,2} \\
\end{smallmatrix}
\right)
\end{equation}
we have 
\begin{equation}
\res A = \left(a_{1,1},a_{1,2},a_{2,1},a_{2,2}\right)^T.
\end{equation}

Using this notation, $NOT$ gate resulting from Alternative
Rule~\ref{qr:states-mtx} is given as
\begin{equation}
\left(
\begin{smallmatrix}
 0 & 0 & 0 & 1 \\
 0 & 0 & 1 & 0 \\
 0 & 1 & 0 & 0 \\
 1 & 0 & 0 & 0 \\
\end{smallmatrix}
\right),
\end{equation}
As expected, the resulting gate is equivalent to the $SWAP$ operation defined
for qubits.

Without using $\res$ mapping, this can be represented as a tensor (\ie\ $n$-way
array),
\begin{equation}
NOT_{\MC{2}} = \left(
\left(
\begin{smallmatrix}
 0 & 0 \\
 0 & 1 \\
\end{smallmatrix}
\right),
\left(
\begin{smallmatrix}
 0 & 0 \\
 1 & 0 \\
\end{smallmatrix}
\right),
 \left(
\begin{smallmatrix}
 0 & 1 \\
 0 & 0 \\
\end{smallmatrix}
\right), 
\left(
\begin{smallmatrix}
 1 & 0 \\
 0 & 0 \\
\end{smallmatrix}
\right) 
\right)
\end{equation}
of dimension $(2,2,4)$. Given a pure state obtained using Alternative Rule
\ref{qr:states-mtx} 
\begin{equation}
X=\left(\begin{smallmatrix}
x_{1,1} & x_{1,2} \\
x_{2,1} & x_{2,2} \\
\end{smallmatrix}\right),
\end{equation}
tensor $NOT_{\MC{2}}$ acts on it as
\begin{equation}
NOT_{\MC{4}} X = (\tr[(NOT_{\MC{2}})_{::1} X],\dots,\tr[(NOT_{\MC{2}})_{::4} X])^{T_{(1,2);3}},
\end{equation}
where $(NOT_{\MC{2}})_{::i}$ denotes $i$-th element in third way (dimension) of
the tensor and $T_{{(1,2)};3}$ is a general transposition operation, which
exchanges between indices $(1,2)$ and $3$.

As a matter of fact, any mapping from bits to $\MC{2}$, \eg\ given as
\begin{equation}
0\mapsto \Span{\left(
\begin{smallmatrix}
1 & 0\\
0 & 1
\end{smallmatrix}
\right), 
\left(
\begin{smallmatrix}
0 & 1\\
1 & 0
\end{smallmatrix}
\right)
},
1\mapsto \Span{\left(
\begin{smallmatrix}
0 & -i\\
i & 0
\end{smallmatrix}
\right), 
\left(
\begin{smallmatrix}
1 & 0\\
0 & -1
\end{smallmatrix}
\right)
}.
\end{equation}
will be isomorphic to quantization obtained using qutrits. The only difference
we get in this case is the representation of quantum operations in the form of
tensors.

This situation shows the importance of the simplicity axiom stated in
\cite{hardy01quantum}. This axiom basically states, that the number of degrees
of freedom should take the minimal value which still enables it to be consistent
with the formulation of the theory. From the above considerations, one can see
the introduced rules, which are based on objects like matrices, do not introduce
any qualitative changes into the quantum theory obtained with them.

\section{Summary}\label{sec:final}
The main goal of this report was to provide a clear and consistent introduction
to the basic concepts of quantum information theory. We aimed at show that the
basic rule of linearity allows an easy \emph{quantization} of classical
information theory and introduction of the concepts of superposition and
entanglement.

Our intension was not to derive quantum theory from a fixed, minimalistic, set
of axioms. Such attempts were made in several papers and the reader interested
in this matter is advised to consult \cite{hardy01quantum, fuchs02quantum,
dariano06operational, fivel12derivation} for the recent developments.

Standard quantum mechanics provides one of possible theories constructed as a
generalization of the classical theory. The study of different approaches for
construction of such theories was undertaken by many
authors~\cite{mielnik74generalized, zyczkowski08quartic,pawlowski12hyperbits}.
The rules introduced in this report allowed the introduction of alternative
methods for obtaining quantization for classical information theory. The
presented alternative rules enable a better understanding of the choices made
during the representation used in quantum information theory, especially in for
the sake the simplicity criterion.

\section{Acknowledgements}
The Author would like to thank K.~\.Zyczkowski for interesting discussions
concerning possible generalizations of quantum theory and R.~Winiarczyk for
encouraging work on this report. Part of this report has been presented as a
talk given in January 2012 in the Institute of Theoretical and Applied
Informatics in Gliwice organized with cooperation with the Faculty of Computer
Science of the Silesian University of Technology. 

The Author acknowledges support by the Polish Ministry of Science and Higher
Education under the grant number IP2011 036371. 

\bibliographystyle{plain}
\bibliography{zeroth_quantization}

\end{document}